\documentclass[preprintnumbers,amsmath,amssymb,prd,twocolumn,showpacs]{revtex4}

\begin{document}

\title{Karlhede's invariant and the black hole firewall proposal}

\author{J. W. Moffat$^{\dag\star}$ and V. T. Toth$^{\dag}$\\~\\}

\affiliation{$^\dag$Perimeter Institute, 31 Caroline St North, Waterloo, Ontario N2L 2Y5, Canada}
\affiliation{$^\star$Department of Physics, University of Waterloo, Waterloo, Ontario N2L 3G1, Canada}

\begin{abstract}
\noindent
The Karlhede invariant is formed from the contraction of the covariant derivative of the Riemann tensor. It is a coordinate invariant that vanishes at the Schwarzschild event horizon $r=2m$. The vanishing of the invariant allows an observer to construct a local measuring device and use it to detect an event horizon while falling into a black hole. Recent proposals postulate the existence of a ``firewall'' at the event horizon that may incinerate an infalling observer. These proposals face an apparent paradox if a freely falling observer detects nothing special in the vicinity of the horizon. The behavior of Karlhede's invariant raises the possibility that the event horizon is a real physical membrane with measurable properties that are detectable by a freely falling observer.
\end{abstract}


\maketitle

It is widely assumed that a freely falling observer cannot determine his distance from the source of a gravitational field using local measurements of the field and no outside knowledge. For instance, while an observer is able to measure tidal forces using a local experiment, in the absence of knowledge of the mass of the source toward which he is falling, tidal force observations are insufficient to determine the distance from that source.

A broadly accepted consequence is that an observer falling toward a black hole singularity will not be able to determine by performing only{\it  local} measurements if, and when, he crossed the horizon.

This is consistent with our understanding that the peculiar properties of the Schwarzschild metric at the event horizon are purely artifacts of the coordinates; no physical singularity exists at the horizon.

Yet more recently, there have been several proposals~\cite{2013JHEP...02..062A} that suggest that the event horizon is a physically special region of spacetime. A resolution of the black hole information loss paradox is offered by positing the existence of a ``firewall'', a region of spacetime at or near the horizon that would incinerate an observer close to the horizon~\cite{2013JHEP...02..062A}.

The firewall proposal is quantum mechanical in nature. However, the possible existence of such a firewall raises renewed interest in the possibility that classical general relativity might also predict unique properties associated with the location of the event horizon.

In 1982, Karlhede et al.,~\cite{1982GReGr..14..569K} investigated the properties of local geometry in terms of the Riemann curvature tensor and its higher derivatives. They discovered that the lowest-order non-trivial scalar term, constructed by contracting the covariant derivative of the curvature tensor with itself, has unexpected properties.

Karlhede's invariant can be written as

\begin{align}
{\cal I}=R^{\alpha\beta\gamma\delta;\kappa}R_{\alpha\beta\gamma\delta;\kappa},
\end{align}
where $R_{\alpha\beta\gamma\delta}$ is the Riemann curvature tensor and the semicolon denotes the covariant derivative with respect to the metric $g_{\mu\nu}$. In Schwarzschild coordinates, represented by the line element
\begin{align}
ds^2=\left(\frac{r-2m}{r}\right)dt^2-\left(\frac{r-2m}{r}\right)^{-1}dr^2-r^2d\Omega^2,
\end{align}
where $d\Omega^2=d\theta^2+\sin^2\theta d\phi^2$, Karlhede's invariant is
\begin{align}
{\cal I}=-\frac{720m^2(r-2m)}{r^9}.\label{eq:I}
\end{align}
From this form, it is immediately evident that ${\cal I}$ changes sign at the horizon when $r=2m$.

Because ${\cal I}$ is a coordinate invariant, Eq.~(\ref{eq:I}) is also true in the comoving Kruskal-Szekeres coordinates of an infalling observer.

The sign change at $r=2m$ explains the significance of this invariant, distinguishing it from other scalar invariants such as the Kretschmann scalar:
\begin{align}
{\cal K}=R^{\alpha\beta\gamma\delta}R_{\alpha\beta\gamma\delta}.
\end{align}
In the Schwarzschild metric, we get
\begin{align}
{\cal K}=\frac{48m^2}{r^6},
\end{align}
which shows no special behavior at the horizon. Even if an observer has the means to measure ${\cal K}$, without prior knowledge about the mass $m$ of the source of the gravitational field, there is no way to determine where the infalling observer is with respect to the horizon. In this respect, measuring ${\cal K}$ is akin to measuring tidal forces; these can also be used to estimate one's position with respect to the horizon, but only if one is in possession of prior knowledge about the mass of the source.

The case of ${\cal I}$ is different. Because it changes sign at the horizon, it can serve as a true ``horizon detector''; if an infalling observer has the means to measure ${\cal I}$, the moment of crossing the horizon can be determined unambiguously, without requiring any prior knowledge about the mass of the source.

Can an observer measure ${\cal I}$? In principle, this measurement involves measuring the components of the curvature tensor and their derivatives. Such a measurement is manifestly possible in a small neighborhood without relying on distant, external observables. For instance, one can envision a device that utilizes different optical paths to measure how the polarization vector of a beam of light changes depending on the path; by performing this measurement at different times and at different locations within a small neighborhood, gradients can also be estimated. (For a different, slightly more elaborate {\em gedankenexperiment}, see \cite{1997GReGr..29..997T}.) Thus, one can imagine a sufficiently sensitive, yet small and compact ``black box'' with a simple numerical or analog display, showing the measured value of ${\cal I}$. When the device crosses the horizon, the display shows zero and changes sign.

The most general case of 
rotating, charged matter is represented by the Kerr-Newman metric:
\begin{align}
ds^2&=\frac{\Delta}{\rho^2}(dt-a\sin^2\theta d\phi)^2\nonumber\\
-&\frac{\sin^2\theta}{\rho^2
}\left[(r^2+a^2)d\phi-adt\right]^2-\rho^2\left(\frac{dr^2}{\Delta}+d\theta^2\right),
\end{align}

where $\rho^2=r^2+a^2\cos^2\theta$ and $\Delta=r^2-2mr+a^2+e^2$. In this metric, ${\cal I}$ can be written in the following form:


\begin{align}
{\cal I}=\frac{-16}{(a^2\cos^2\theta+r^2)^9}\sum\limits_{i=0}^{i=5}T_ia^{2i}\cos^{2i}\theta,
\end{align}
where
\begin{align}
T_0=&[45m^2r^4-18m(5m^2+6e^2)r^3+e^2(261m^2+76e^2)r^2\nonumber\\
&-4me^2(65e^2+9a^2)r+4e^4(19e^2+11a^2)]r^6,\\
T_1=&-[1215m^2r^4-180m(14m^2+11e^2)r^3\nonumber\\
&+4e^2(1341m^2+178e^2)r^2-12e^2(295e^2+27a^2)mr\nonumber\\
&+12e^4(62e^2+13a^2)]r^4,\\
T_2=&6[315m^2r^4-6m(175m^2+37e^2)r^3+1545e^2m^2r^2\nonumber\\
&\hskip -1em +2e^2(15a^2-353e^2)mr+2e^4(49e^2-13a^2)]r^2,\\
T_3=&2[945m^2r^4+90m(14m^2-15e^2)r^3\nonumber\\
&+2e^2(178e^2-585m^2)r^2+2e^2(151e^2-45a^2)mr\nonumber\\
&+2e^4(11a^2-8e^2)],\\
T_4=&-1215m^2r^2-90m(m^2-8e^2)r\nonumber\\
&-e^2(76e^2-45m^2),\\
T_5=&45m^2.
\end{align}

When $a=0$ (Reissner-Nordstr\"om metric), ${\cal I}$ ceases to be dependent on $\theta$ and can be simplified:
\begin{align}
{\cal I}\hskip -0.05em =\hskip -0.15em \frac{-\hskip -0.15em 16(r^2\hskip -0.15em -\hskip -0.15em 2mr\hskip -0.15em +\hskip -0.15em e^2)(45m^2r^2\hskip -0.15em -\hskip -0.15em 108e^2mr\hskip -0.15em +\hskip -0.15em 76e^4)}{r^{12}},
\end{align}
with real roots at $r=m\pm\sqrt{m^2-e^2}$, indicating that ${\cal I}$ changes sign at the event horizon and once again at the internal Cauchy horizon.

If instead we set $e=0$ (Kerr solution), we get
%
\begin{align}
{\cal I}=\frac{-720m^2(a^2\cos^2\theta+r^2-2mr)Q_1Q_2}{(a^2\cos^2\theta+r^2)^9},
\end{align}
where
\begin{align}
Q_1&=(a\cos\theta-r)^4-4ar^2\cos\theta(3a\cos\theta-2r),\\
Q_2&=(a\cos\theta-r)^4-4a^2r\cos^2\theta(3r-2a\cos\theta).
\end{align}

In particular, it is notable that in the case of a rotating Kerr black hole, ${\cal I}$ changes sign not at the horizon, but on the ergosphere:
\begin{equation}
r=m\pm\sqrt{m^2-a^2\cos^2\theta}.
\end{equation}
This is consistent with the finding~\cite{2007CQGra..24.2929S} that ${\cal I}$ cannot be used as a reliable ``horizon detector'' in the non-spherically symmetric case, though it remains to be seen what, if any, physical significance ${\cal I}$ has in these situations.

We note that, as discussed in \cite{2004GReGr..36.1159L}, a further eight solutions in the Kerr case are given by
\begin{align}
r&=\pm(1+\sqrt{2}\pm\sqrt{4+2\sqrt{2}})a\cos\theta,\\
r&=\pm(1-\sqrt{2}\pm\sqrt{4-2\sqrt{2}})a\cos\theta.
\end{align}
Numerically, these eight values correspond to
\begin{equation}
r=\pm (0.199,0.668,1.497,5.027)a\cos\theta.
\end{equation}
These solutions are also discussed in detail in \cite{1998gr.qc.....8055G}.

The fact that the vanishing of Karlhede's invariant ${\cal I}$ at the Schwarzschild event horizon $r=2m$ can be physically measured, based on classical general relativity, as an observer freely falls into a black hole implies the reality of the event horizon as a physical membrane. A possible interpretation of such a membrane is in the form of a firewall (for a recent discussion, see \cite{2014arXiv1403.7470I}). Such a firewall may be experienced by an observer hovering close to a black hole event horizon, who detects a large surface gravity and high temperature Hawking radiation via Tolman's formula for the local temperature:
\begin{align}
T=\frac{T_H}{(g_{00})^{1/2}},
\end{align}
where $T_H$ is the Hawking temperature. Indeed, at the horizon, $r=2m$, the surface gravity is infinite and this would produce an infinite heat bath and Unruh temperature that can be interpreted as a firewall. However, this interpretation creates the paradox that a freely falling observer does not experience any firewall-like effects. On the other hand, the existence of Karlhede's invariant ${\cal I}$ suggests that the horizon has physical significance even for a a freely falling observer. Whether or not this is sufficient justification to consider the potential existence of the firewall as a physically viable proposal requires further investigation.

\acknowledgments

The authors thank Jianwei Mei for helpful comments. JWM thanks the John Templeton Foundation for their generous support of this research.  Research at the Perimeter Institute for Theoretical Physics is supported by the Government of Canada through Industry Canada and by the Province of Ontario through the Ministry of Research and Innovation (MRI). JWM thanks the hospitality of the Institut d'Astrophysique de Paris (IAP) where this research was completed.

\bibliography{refs}

\end{document}